\documentstyle[12pt,axodraw]{article}

\catcode`@=12
\begin{document}
\thispagestyle{empty}
\begin{flushright} 
UCRHEP-T305\\ 
April 2001\
\end{flushright}
\vspace{0.5in}
\begin{center}
{\large	\bf Quark Mass Matrices from a Softly Broken U(1) Symmetry\\}
\vspace{1.5in}
{\bf Ernest Ma\\}
\vspace{0.2in}
{\sl Physics Department, University of California, Riverside, 
California 92521\\}
\vspace{1.5in}
\end{center}
\begin{abstract}\
Assigning U(1) charges to the quarks of the standard model, and 
allowing one extra scalar doublet with $m^2 > 0$, the correct pattern of the 
$up$ and $down$ quark mass matrices is obtained, together with their 
charged-current mixing matrix.
\end{abstract}
\newpage
\baselineskip 24pt
In the standard model of particle interactions, quark masses and the 
charged-current mixing matrix, $V_{CKM}$, which links the $(d,s,b)_L$ 
quarks to the $(u,c,t)_L$ quarks, are known to exhibit a hierarchical 
pattern \cite{pdg}.
\begin{eqnarray}
&& m_u \sim 1-5~{\rm MeV}, ~~~ m_d \sim 3-9~{\rm MeV}, ~~~ m_s \sim 75-170 
~{\rm MeV}, \nonumber \\ && m_c \sim 1.15-1.35~{\rm GeV}, ~~~ m_b \sim 
4.0-4.4~{\rm GeV}, ~~~ m_t = 174.3 \pm 5.1~{\rm GeV},
\end{eqnarray}
and
\begin{equation}
V_{CKM} = \left[ \begin{array} {c@{\quad}c@{\quad}c} 0.9742-0.9757 & 
0.219-0.226 & 0.002-0.005 \\ 0.219-0.225 & 0.9734-0.9749 & 0.037-0.043 \\ 
0.004-0.014 & 0.035-0.043 & 0.9990-0.9993 \end{array} \right],
\end{equation}
where the magnitude range of each matrix element is denoted.

With the one Higgs doublet of the standard model, this pattern (or any other) 
is certainly allowed, but then Yukawa couplings spanning 5 decades of 
magnitude are needed.  On the other hand, if two Higgs doublets exist with 
$v_1 = 174$ GeV, but $v_2/v_1 \sim 10^{-3} \sim 10^2$ MeV, then Yukawa 
couplings spanning only 2 decades of magnitude are sufficient.  In other 
words, $m_{c,b,t}$ are proportional to $v_1$, but $m_{u,d,s}$ are proportional 
to $v_2$.  Of course, the hierarchical structure of the 2 vacuum expectation 
values (VEVs) is yet to be explained.  As shown below, this may be attributed 
to the soft breaking of an assumed U(1) symmetry and is easily implemented 
if $\Phi_2$ has $m^2 > 0$ while $\Phi_1$ has $m^2 < 0$ as in the standard 
model.

The puzzle of quark masses and the charged-current mixing matrix, usually 
denoted by $V_{ij}$, with $i=u,c,t$ and $j=d,s,b$, has received a great 
deal of continuing attention.  One approach is to restrict the number of 
independent parameters necessary for a general description of all masses 
and mixing angles, so that a relationship among them may be derived, such 
as \cite{weinberg}
\begin{equation}
V_{us} = \sqrt {m_d \over m_s}.
\end{equation}
This is usually postulated without recourse to a well-defined symmetry of 
the Lagrangian \underline {nor} the extra particle content required to 
sustain it \cite{except}.  Another shortcoming of this approach is that 
the mass hierarchy of Eq.~(1) remains largely unexplained.

The present approach is different.  It looks for a way to understand why 
$m_{u,d,s} << v = 174$ GeV, i.e. the scale of electroweak symmetry breaking, 
as well as the pattern of Eq.~(2).  However, no precise prediction 
such as Eq.~(3) will be made.  This approach was used in a radiative scheme 
some years ago \cite{ma90}, but the model itself is rather complicated.
In contrast, the model to be described below is much simpler, requiring 
only one extra Higgs doublet together with a softly broken global U(1) 
symmetry.

The U(1) assignments of the 3 generations of quarks and the 2 Higgs doublets 
are given as follows.
\begin{eqnarray}
&& (u,d)_L \sim 1, ~~ u_R \sim 2, ~~ d_R \sim 0; \\ 
&& (c,s)_L \sim 1, ~~ c_R \sim 1, ~~ s_R \sim 0; \\
&& (t,b)_L \sim 0, ~~ t_R \sim 0, ~~ b_R \sim 0; \\
&& (\phi_1^+,\phi_1^0) \sim 0, ~~ (\phi_2^+,\phi_2^0) \sim 1.
\end{eqnarray}
As a result, the $up$ quark mass matrix linking $\overline {(u,c,t)}_L$ to 
$(u,c,t)_R$ is given by
\begin{equation}
{\cal M}_u = \left[ \begin{array} {c@{\quad}c@{\quad}c} f_u v_2 & 0 & 0 \\ 
f_{cu} v_2 & f_c v_1 & 0 \\ 0 & f_{tc} v_2 & f_t v_1 \end{array} \right],
\end{equation}
where $v_i = \langle \phi_i^0 \rangle$, and the freedom to rotate among 
$(u,d)_L$ and $(c,s)_L$ has been used to set the $\bar u_L c_R$ element 
to zero; whereas the $down$ quark mass matrix linking $\overline {(d,s,b)}_L$ 
to $(d,s,b)_R$ is given by
\begin{equation}
{\cal M}_d = \left[ \begin{array} {c@{\quad}c@{\quad}c} f_d v_2 & f_{ds} v_2 
& f_{db} v_2 \\ 0 & f_s v_2 & f_{sb} v_2 \\ 0 & 0 & f_b v_1 \end{array} 
\right],
\end{equation}
where the freedom to rotate among the $(d,s,b)_R$ states has been used to 
set the 3 lower off-diagonal entries to zero. 

Assuming $v_2 << v_1$, as well as $f_d \sim f_{ds} \sim f_{db}$ and 
$f_s \sim f_{sb}$, then
\begin{eqnarray}
&& m_u = f_u v_2, ~~ m_d = f_d v_2, ~~ m_s = f_s v_2; \\ 
&& m_c = f_c v_1, ~~ m_b = f_b v_1, ~~ m_t = f_t v_1.
\end{eqnarray}
As for $V_{CKM}$, the contribution from ${\cal M}_u$ is negligible because 
they are of order $(m_u/m_c^2) f_{cu} v_2$ and $(m_c/m_t^2) f_{tc} v_2$. 
Hence
\begin{eqnarray}
&& V_{cb} \simeq {f_{sb} v_2 \over f_b v_1} \simeq 
{f_{sb} \over f_s} {m_s \over m_b} \simeq {f_{sb} \over f_s} (0.017 - 0.043), 
\\ && V_{ub} \simeq {f_{db} v_2 \over f_b v_1} \simeq 
{f_{db} \over f_d} {m_d \over m_b} \simeq {f_{db} \over f_d} (0.001 - 0.002), 
\\ && V_{us} \simeq {f_{ds} v_2 \over f_s v_2} \simeq 
{f_{ds} \over f_d} {m_d \over m_s} \simeq {f_{ds} \over f_d} (0.02 - 0.12).
\end{eqnarray}
Comparing the above with Eq.~(2), it is also clear that the Yukawa coupling 
ratios $f_{ds}/f_d$, $f_{db}/f_d$, and $f_{sb}/f_s$ may all be of order 
unity.  Thus the correct pattern of quark masses and mixing angles is 
obtained.  Obviously, the charged-lepton masses may be treated in the same 
way, i.e.
\begin{equation}
m_e = f_e v_2, ~~ m_\mu = f_\mu v_2, ~~ m_\tau = f_\tau v_1.
\end{equation}
What remains to be shown is how $v_2 << v_1$ can arise naturally.

The most general scalar potential of the 2 assumed scalar doublets is 
given by
\begin{eqnarray}
V &=& m_1^2 \Phi_1^\dagger \Phi_1 + m_2^2 \Phi_2^\dagger \Phi_2 + 
{1 \over 2} \lambda_1 (\Phi_1^\dagger \Phi_1)^2 + {1 \over 2} \lambda_2 
(\Phi_2^\dagger \Phi_2)^2 \nonumber \\ &+& \lambda_3 (\Phi_1^\dagger \Phi_1)
(\Phi_2^\dagger \Phi_2) + \lambda_4 (\Phi_1^\dagger \Phi_2)(\Phi_2^\dagger 
\Phi_1) + [\mu_{12}^2 \Phi_1^\dagger \Phi_2 + h.c.],
\end{eqnarray}
where the $\mu_{12}^2$ term breaks the U(1) symmetry softly.  The equations 
of constraint for the VEVs are then
\begin{eqnarray}
&& v_1[m_1^2 + \lambda_1 v_1^2 + (\lambda_3 + \lambda_4) v_2^2] + \mu_{12}^2 
v_2 = 0 \\ && v_2[m_2^2 + \lambda_2 v_2^2 + (\lambda_3 + \lambda_4) v_1^2] 
+ \mu_{12}^2 v_1 = 0.
\end{eqnarray}
Let $m_1^2 < 0$, $m_2^2 > 0$, and $|\mu_{12}^2| << m_{2}^2$, 
then
\begin{eqnarray}
v_1^2 &\simeq& - {m_1^2 \over \lambda_1}, \\ 
v_2 &\simeq& {-\mu_{12}^2 v_1 \over m_2^2 + (\lambda_3 + \lambda_4) v_1^2}.
\end{eqnarray}
Since the $\mu_{12}^2$ term breaks the U(1) symmetry, it is natural \cite{th}
for it to be small compared to $m_{2}^2$.  Thus
\begin{equation}
v_2 << v_1
\end{equation}
is obtained.

The physical scalar sector of this model consists of a standard-model-like 
neutral Higgs boson $H$ (which is mostly $Re \phi_1^0$) and a heavy doublet 
of mass $m_2$ approximately.  The dominant decays of $H$ are the same as 
in the standard model, i.e. into $\bar t t$, $Z Z$, $W^+ W^-$, $\bar b b$, 
$\bar c c$, and $\tau^+ \tau^-$.  However, its decays into other final states 
are modified because they depend on the mixing of $\Phi_2$ with $\Phi_1$. 
In practice, it will be very difficult to tell the difference because the 
latter decay modes are very much suppressed.

From Eqs.~(8) and (9), it is clear that there are flavor-changing 
neutral-current (FCNC) interactions in this model, but they are suitably 
suppressed, as explained below.  The matrices ${\cal M}_u$ of Eq.~(8) and 
${\cal M}_d$ of Eq.~(9) are diagonalized according to
\begin{eqnarray}
V_u^\dagger {\cal M}_u U_u &=& \left( \begin{array} {c@{\quad}c@{\quad}c} 
m_u & 0 & 0 \\ 0 & m_c & 0 \\ 0 & 0 & m_t \end{array} \right), \\ 
V_d^\dagger {\cal M}_d U_d &=& \left( \begin{array} {c@{\quad}c@{\quad}c} 
m_d & 0 & 0 \\ 0 & m_s & 0 \\ 0 & 0 & m_b \end{array} \right),
\end{eqnarray}
where
\begin{equation}
V_{CKM} = V_u^\dagger V_d,
\end{equation}
but since $V_u = 1$ to a very good approximation, $V_{CKM} \simeq V_d$, and 
the $(d,s,b)_L$ states have to be rotated by $V_d$ to become mass eigenstates. 
For example, $b_L$ in Eq.~(9) becomes $V_{ub}^* d_L + V_{cb}^* s_L + 
V_{tb}^* b_L$ in the mass-eigenstate basis.  Similarly, the $(d,s,b)_R$ 
states are rotated by $U_d$, i.e.
\begin{equation}
U_d \simeq \left[ \begin{array} {c@{\quad}c@{\quad}c} V_{ud} & (m_d/m_s) 
V_{us} & (m_d/m_b) V_{ub} \\ (m_d/m_s) V_{cd} & V_{cs} & (m_s/m_b) V_{cb} \\ 
(m_d/m_b) V_{td} & (m_s/m_b) V_{ts} & V_{tb} \end{array} \right].
\end{equation}
Thus $b_R$ becomes $(m_d/m_b) V_{ub}^* d_R + (m_s/m_b) V_{cb}^* s_R + 
V_{tb}^* b_R$ in the mass-eigenstate basis.

In the $up$ quark sector, the roles of $V$ and $U$ are reversed, i.e.
\begin{equation}
U_u \simeq \left[ \begin{array} {c@{\quad}c@{\quad}c} 1 & f_{cu} v_2/m_c & 0 
\\ -f_{cu} v_2/m_c & 1 & f_{tc} v_2/m_t \\ 0 & -f_{tc} v_2/m_t & 1 \end{array} 
\right],
\end{equation}
and
\begin{equation}
V_u \simeq \left[ \begin{array} {c@{\quad}c@{\quad}c} 1 & f_{cu} v_2 m_u/m_c^2 
& 0 \\ -f_{cu} v_2 m_u/m_c^2 & 1 & f_{tc} v_2 m_c/m_t^2 \\ 0 & -f_{tc} v_2 
m_c/m_t^2 & 1 \end{array} \right],
\end{equation}
which is the identity matrix to a very good approximation, as mentioned 
earlier.  Thus $c_L$ becomes $-(m_u/m_c^2) f_{cu} v_2 u_L + c_L + (m_c/m_t^2) 
f_{tc} v_2 t_L$ and $c_R$ becomes $-f_{cu} (v_2/m_c) u_R + c_R + f_{tc} 
(v_2/m_t) t_R$ in the mass-eigenstate basis.

Consider now the phenomenology of the $down$ quark sector.  Since $\bar b_L 
b_R$ is the only term which couples to $\Phi_1$, if it is replaced by 
$\Phi_2$, there would be no FCNC interactions at all in this sector.  Hence 
all FCNC effects are contained in the term $f_b \bar b_L b_R [\phi_1^0 - 
(v_1/v_2) \phi_2^0] + h.c.$, i.e.
\begin{eqnarray}
&& f_b [ V_{ub} V_{tb}^* \bar d_L b_R + V_{cb} V_{tb}^* \bar s_L b_R + 
V_{ub} V_{cb}^* (m_s/m_b) \bar d_L s_R + V_{tb} V_{cb}^* (m_s/m_b) 
\bar b_L s_R \nonumber \\ && + V_{cb} V_{ub}^* (m_d/m_b) \bar s_L d_R + V_{tb} 
V_{ub}^* (m_d/m_b) \bar b_L d_R] [\phi_1^0 - (v_1/v_2) \phi_2^0] + h.c.
\end{eqnarray}
in the mass-eigenstate basis.  The most severe constraint on $m_2$ comes 
from the $b \to s \mu^+ \mu^-$ rate through $\phi_2^0$ exchange, i.e.
\begin{equation}
{\Gamma (b \to s \mu^+ \mu^-) \over \Gamma (b \to c \bar \nu_\mu \mu^-)} 
\simeq {f_b^2 f_\mu^2 v_1^2 \over 32 G_F^2 m_2^4 v_2^2} < {5.2 \times 10^{-6} 
\over 0.102} = 5.1 \times 10^{-5},
\end{equation}
where the experimental $B^+ \to K^+ \mu^+ \mu^-$ upper bound has been used 
for $b \to s \mu^+ \mu^-$, which is of course an overestimate.  In other 
words, the numerical bound of Eq.~(29) may not be as small in reality.  
Using $f_b = m_b/v_1 = 4.2/174 = 0.024$ and $f_\mu = m_\mu/v_2$, Eq.~(29) 
implies
\begin{equation}
m_2 v_2 > 968 ~{\rm GeV}^2.
\end{equation}
Thus $v_2 = 200$ MeV requires $m_2 > 4.84$ TeV.

The $K_L - K_S$ mass difference $\Delta m_K$ gets its main contribution 
from $(\bar d_L s_R)(\bar d_R s_L)$ in this model through $\phi_2^0$ 
exchange.  Thus
\begin{equation}
{\Delta m_K \over m_K} \simeq {B_K f_K^2 v_1^2 \over 3 m_2^2 v_2^2} f_b^2 
|V_{ub} V_{cb}|^2 {m_s m_d \over m_b^2}.
\end{equation}
Using $f_K = 114$ MeV, $B_K = 0.4$, $|V_{ub}| = 0.0035$, $|V_{cb}| = 0.040$, 
$m_s = 125$ MeV, $m_d = 7$ MeV, and Eq.~(30), this contribution is then less 
than $3.1 \times 10^{-20}$, which is certainly negligible compared against 
the experimental value of $7.0 \times 10^{-15}$.

Similarly, the $\Delta m_{B^0}$ and $\Delta m_{B_s^0}$ contributions are
\begin{equation}
{\Delta m_{B^0} \over m_{B^0}} \simeq {B_B f_B^2 v_1^2 \over 3 m_2^2 v_2^2} 
f_b^2 |V_{ub} V_{tb}|^2 {m_d \over m_b},
\end{equation}
and
\begin{equation}
{\Delta m_{B_s^0} \over m_{B_s^0}} \simeq {B_B f_B^2 v_1^2 \over 3 m_2^2 
v_2^2} f_b^2 |V_{cb} V_{tb}|^2 {m_s \over m_b}.
\end{equation}
Using $f_B = 170$ MeV, $B_B = 1.0$, $|V_{tb}| = 1$, and the other parameter 
values as before, these contributions are respectively less than $3.7 \times 
10^{-15}$ and $8.5 \times 10^{-12}$, to be compared against the experimental 
value of $5.9 \times 10^{-14}$ for the former and the experimental lower 
bound of $1.3 \times 10^{-12}$ for the latter.  

In the case of $D^0 - \overline {D}^0$ mixing, the main contribution comes 
from $(\bar c_L u_R)(\bar c_R u_L)$, i.e.
\begin{equation}
{\Delta m_{D^0} \over m_{D^0}} \simeq {B_D f_D^2 v_1^2 \over 3 m_2^2} f_c^2 
f_{cu}^2 {m_u \over m_c^3}.
\end{equation}
Using $f_D = 150$ MeV, $B_D = 0.8$, $m_c = 1.25$ GeV, $f_c = f_{cu} = m_c/v_1 
= 0.0072$, and $m_u = 4$ MeV, this contribution is then $1.0 \times 10^{-15}$ 
(1 TeV/$m_2)^2$, well below the experimental upper bound of $2.5 \times 
10^{-14}$.

Other FCNC processes are also suppressed. For example,
\begin{equation}
\Gamma (K_L \to \mu^+ \mu^-) \simeq {f_K^2 m_K^3 \over 64 \pi} {f_b^2 m_\mu^2 
v_1^2 \over m_2^4 v_2^4} |V_{ub} V_{cb}|^2 {m_s^2 \over m_b^2}.
\end{equation}
Using the previously chosen values for all the parameters, this contribution 
is less than $3.1 \times 10^{-29}$, well below the experimental value of 
$9.2 \times 10^{-26}$ GeV.  As for $K_L \to e^+ e^-$, it is further 
suppressed by $m_e^2/m_\mu^2$, resulting in a contribution of less than 
$7.2 \times 10^{-34}$, which is even more negligible compared to the 
experimental value of $1.1 \times 10^{-28}$ GeV.  Finally, the $b \to s 
\gamma$ rate receives a contribution proportional to $|f_b V_{cb}|^2/m_2^2$, 
which would be competitive with the standard model if $f_b$ were of order 
unity and $m_2$ of order $M_W$, but since $f_b = 0.024$ and $m_2 >> M_W$ in 
this model, it is again negligible.

There is also a contribution from $\Phi_2$ to the muon anomalous magnetic 
moment \cite{g-2}.  It is easily calculated to be
\begin{equation}
\Delta a_\mu = {f_\mu^2 \over 16 \pi^2} {m_\mu^2 \over m_2^2} \left( 
\ln {m_2^2 \over m_\mu^2} - 1 \right),
\end{equation}
which is of the order $10^{-11}$ or less, and thus negligible.  However, 
the present model can be extended to allow for neutrino masses using a 
leptonic Higgs doublet \cite{ma01}, then the possible observed discrepancy 
in $\Delta a_\mu$ may be explained \cite{mr}, but a nearly degenerate 
neutrino mass matrix is required.  The extra contributions from $\Phi_2$ to 
the oblique parameters $S,T,U$ in precision electroweak measurements are all 
suppressed by $\lambda_4 v_1^2/m_2^2$ and do not upset the excellent fit of 
the standard model.

In summary, a new realization of the generation structure of quarks and 
leptons has been presented in this paper, as given by Eqs.~(4) to (7). 
The one extra scalar doublet is heavy with $m^2 > 0$.  Typical values are 
$m_2 \sim$ few TeV with $v_2 \sim$ fraction of a GeV, whereas $\Phi_1$ has 
$m^2 < 0$, resulting in $v_1 = 174$ GeV and $m_H = 115$ GeV or greater.  
This is accomplished by a softly broken U(1) symmetry with $|\mu_{12}|^2
/m_2^2 \sim 10^{-3}$.  The pattern of the observed quark masses (with 
$m_{u,d,s}$ from $v_2$ and $m_{c,b,t}$ from $v_1$) and the corresponding 
charged-current mixing matrix ($V_{CKM}$) is realized without severely 
hierarchical Yukawa couplings.  All FCNC effects of this model are suitably 
suppressed if $m_2 v_2 > 968$ GeV$^2$ and do not change the good agreement 
of the standard model with present data.  The standard-model-like neutral 
Higgs boson of this model has dominant decays identical to those of the 
standard model.  To distinguish the two models, the discovery of $b \to s 
\mu^+ \mu^-$ would help, but finding the extra Higgs doublet $\Phi_2$ would 
be more decisive.  If $m_2$ is nearer 1 TeV than 5 TeV, then future 
high-energy accelerators such as the Large Hadron Collider (LHC) will have 
a reasonable chance of doing it.

This work was supported in part by the U.~S.~Department of Energy
under Grant No.~DE-FG03-94ER40837.

\newpage
\bibliographystyle{unsrt}

\end{document}